\documentclass{article}

\usepackage{PRIMEarxiv}

\usepackage[utf8]{inputenc} 
\usepackage[T1]{fontenc}    
\usepackage{hyperref}       
\usepackage{url}            
\usepackage{booktabs}       
\usepackage{amsfonts}       
\usepackage{nicefrac}       
\usepackage{microtype}      
\usepackage{lipsum}
\usepackage{fancyhdr}       
\usepackage{graphicx}       
\graphicspath{{media/}}     
\usepackage[ruled]{algorithm2e} 

\title{PRIME: A Price-Reverting Impact Model of a cryptocurrency Exchange
}

\author{
  Christopher J. Cho \\
  School of Electronics and Computer Science \\
  University of Southampton \\
  United Kingdom\\
  \texttt{c.j.cho@soton.ac.uk} \\
   \And
  Timothy J. Norman \\
  School of Electronics and Computer Science \\
  University of Southampton \\
  United Kingdom\\
  \texttt{t.j.norman@soton.ac.uk} \\
   \And
  Manuel Nunes \\
  Southampton Business School \\
  University of Southampton \\
  United Kingdom\\
  \texttt{manuel.nunes@soton.ac.uk} \\
}

\begin{document}
\maketitle

\begin{abstract}
In a financial exchange, market impact is a measure of the price change of an asset following a transaction. This is an important element of market microstructure, which determines the behaviour of the market following a trade. In this paper, we first provide a discussion on the market impact observed in the BTC/USD Futures market, then we present a novel multi-agent market simulation that can follow an underlying price series, whilst maintaining the ability to reproduce the market impact observed in the market in an explainable manner. This simulation of the financial exchange allows the model to interact realistically with market participants, helping its users better estimate market slippage as well as the knock-on consequences of their market actions. In turn, it allows various stakeholders such as industrial practitioners, governments and regulators to test their market hypotheses, without deploying capital or destabilising the system.
\end{abstract}


\section{Introduction}
A market simulator is an environment that allows users to test their market strategies in a synthetic setting. These simulators are useful, as it enables the participants to gather feedback on their ideas without deploying capital into the markets. Consequently, many researchers and institutions commit resources to create a realistic simulation of financial exchanges. While a lot of work remains behind closed doors due to the proprietary nature of this research, we have seen an increased number of open-source research in this domain over the past few years \cite{DBLP:journals/corr/abs-1904-12066, vyetrenko2019get, cho2021bit, coletta2022learning}.

Most of these past works look at the exchange with a focus on the market microstructure from a top-down perspective, looking at the behaviour of the system as a whole. However, the bottom-up analysis is yet to be studied extensively, especially from a transaction-by-transaction basis. This approach is key to understanding the evolution of the limit order book over time. In particular, research into simulations to reproduce realistic market impact remains largely untouched, despite many studies looking at analysing and modelling observed market impact in the real world in a theoretical manner. This poses a problem, as a realistic market simulation should interact with market participants in the presence of market impact and slippage. We address this gap in research by offering two key contributions. First, we identify and analyse the components of market microstructure observed in the BTC/USD Futures market to be used as a benchmark for our simulation. Secondly, we present a novel simulation framework, that leverages and enhances an existing model, which produces desirable market microstructural dynamics in an explainable manner.

Our contributions will have various implications for stakeholders in the world of market simulations. Industrial practitioners will be able to test the performance of their strategies without deploying capital into the market. On the other hand, governments and regulators will be able to run stress tests and scenario analyses of the markets, without intervening or destabilising the real-world exchange. Lastly, researchers within academia can benefit from a realistic model of an exchange to test, train and benchmark their theories against others, in a controlled and reproducible manner.

\section{Background Concepts}

To understand the simulation of financial exchange and its usage, a basic understanding of the exchange, as well as common methods to analyse and simulate the exchange needs to be understood.

\subsection{Continuous Double Auction}

The financial exchange is a Continuous Double Auction (CDA), where multiple buyers and sellers simultaneously transact a homogeneous product in a centralised venue. These participants trade with one another by using a standardised order format that contains information on price, volume, and the buy/sell status. Additionally, the orders are divided into two different types: limit orders (LO) and market orders (MO). Limit orders can be viewed as the participant's intention to trade. These orders are placed inside the limit order book (LOB) at the price and quantity specified by the agent and wait for a counterparty to the trade to appear. On the other hand, market orders are transacted immediately against the current best limit order, by becoming the counterparty to an existing limit order.

\subsection{Limit Order Book}

The Limit Order Book is a histogram of limit order volume at each discrete price level of a given exchange. This histogram evolves over time, as limit orders are added and taken out of the order book via market orders and cancellations. This means that in theory, every outstanding order in the order book can be mapped to a participant, alongside the timestamp of the order. However, most exchanges limit this information by anonymising the orders to protect the identity of their clients. In addition, there are specific terminologies used in the Limit Order Book. A ``bid'' is a limit order looking for a counterparty to buy the asset from, whereas an ``ask'' is a limit order looking to sell the asset. At any point in time, the ``spread'' of the market is defined as the gap between the best bid and ask of the market, and the ``mid-price'' is defined as the halfway point between the best bid and ask price of the market. 

\subsection{Market Simulator}

In order to create a realistic market simulation, a simulator that can accommodate our model is necessary. Out of a handful of open-source market simulators, we take the ABIDES \cite{DBLP:journals/corr/abs-1904-12066} (the \textbf{A}gent-\textbf{B}ased \textbf{I}nteractive \textbf{D}iscrete \textbf{E}vent \textbf{S}imulation environment) platform and add additional utilities to accommodate our simulation. This platform was chosen over others for a few reasons. Firstly, it has a built-in script that allows for easy parallelisation of the simulation workflow, allowing us to carry out large-scale empirical studies with less effort. Secondly, the environment has built-in examples of the ``agent'' class, that include simple, but useful and commonly used agents such as \textit{Zero Intelligence} and \textit{Noise Trader}. Lastly, the platform allows participating agents to query an ``oracle'', that can provide the querying agents with information that is exogenous to the simulation (e.g. true price of the asset using a historic dataset). This last point is crucial, as it opens up the possibilities for the simulation to track the historic price series of an asset, which allows for a better comparison between historic and simulated market data.

\section{Data Collection}

Financial assets are traded either on an exchange or off the exchange in an ``over the counter'' (OTC) fashion. This research focuses on the former, as OTC transactions are done privately with little information shared with the public. Data from exchanges are in general homogeneous, well-maintained, and plentiful. However, exchanges of traditional assets such as equities and futures charge users to access their order book data, which is a huge barrier to entry for academic researchers. On the other hand, data from the less established cryptocurrency exchange comes free of charge. Therefore, our research is carried out on the Level 3 LOB data collected from the Binance API. More specifically, we queried data from the USD/BTC perpetual futures market from July 2020 to October 2021. This is an order-by-order breakdown of all executed orders during this period. Unfortunately, data is unavailable for limit order submissions and cancellations, somewhat limiting the scope of our research. In addition, we recognise that due to the decentralised nature of cryptocurrencies, the data from Binance may not be exactly in-line with the entire cryptocurrency market. However, due to their size (largest exchange by transaction volume, March 2023) and the existence of high-frequency traders who arbitrage between various cryptocurrency exchanges, we assume that the data we collect from Binance is representative of the entire market for the purpose of this research.

\section{Market Impact in the Real World}
\label{Section:market_impact_in_the_real_world}

\subsection{What is Market Impact?}
\label{Subsection:what_is_market_impact}

Market impact is defined as the change in the price of an asset caused by the trading of that asset. From the perspective of a single order, the market impact is deterministic based on the size and type of the order, and the prevailing liquidity in the limit order book at the point in time of the order. This means the price move is larger if the traded volume is large or prevailing liquidity is thin. In addition, the available liquidity at each price increases roughly linearly as the price moves further away from the best bid/ask price (until a certain distance). This leads to the phenomenon where price requires quadratically increasing trade volume to move linearly away from the mid-price. This is one of the fundamental rational behind the \textit{Square Root law} \cite{torre1997barra} of market impact. 

The market impact becomes less mechanical when approached from a non-instantaneous perspective. This is due to the concept of impact reversion and the hidden liquidity arising from what is commonly known as the \textit{Latent Order Book} \cite{mahanti2008latent}. This is the idea that most market participants do not place limit orders at their private valuation until their valuation becomes (or close to) the most competitive price, in order to increase their likelihood of a transaction. This allows participants to withhold private information from the market, allowing them to better utilise their information advantage. For example, an individual willing to sell BTC at USD 100 may not place their limit order when the best ask price in the market is at USD 98. However, when a large buy order eats up all the liquidity between USD 98 and USD 100, pushing up the best ask price to USD 101, this individual will place a limit order asking for USD 100. This behaviour is encouraged, as placing a large limit order at USD 100 when the market is at USD 98 would reveal to the market that a large seller is in play, thus influencing participants' decisions (e.g. would-be buyers may be more inclined to wait until the price falls, rather than transacting immediately). Consequently, it is only when the price level changes, the hidden liquidity is brought out into the market, and causes a reversion in market impact over time. The \textit{Propagator model} \cite{gatheral2010no} is a widely used model that explains the observed part of this phenomenon by dividing up the market impact into initial impacts and the reversion of these market impacts when estimating the total market impact over the current period. We too will use this framework presented by Gatheral et al for the purpose of our research. Note that modelling the idea of visible and latent liquidity in a fully consistent model remains an active area of research \cite{donier2015fully}, and is beyond the scope of our project. Instead, our work focuses on the proof of concept of a model that can replicate the general behaviour of the simulated markets to be in line with the real world in a qualitative manner, from the lens of the \textit{Propagator model}. Thus we leave the exact empirical tuning and mathematical derivations of the model to future work.

\subsection{Initial Market Impact}

As discussed above in Section~\ref{Subsection:what_is_market_impact}, the market impact should be approached from a temporal perspective to avoid drawing conclusions based purely on the mechanical nature of the LOB. Therefore we resample the trade data into 5-second time windows, where the traded volume is defined as the net buy/sell volume across the interval (e.g. 200 buy orders and 500 sell orders within the interval will register as -300 orders), and price change is the difference between the close and open mid price of the 5-second interval.

\begin{figure}[!htb]
  \centering
  \includegraphics[width=\textwidth]{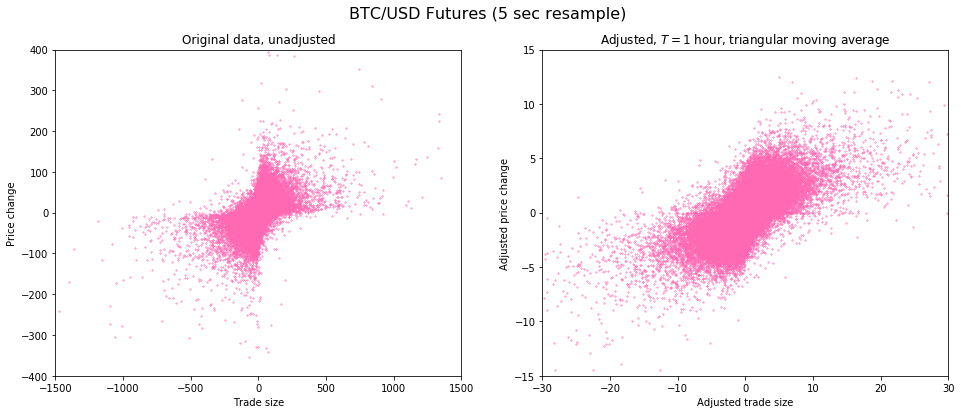}
  \caption{BTC/USD Futures Market Impact (5 sec resample)}
  \label{Figure:BTCUSD_market_impact_5s_resample}
\end{figure}

Looking at the resampled data from the BTC/USD Futures market (Figure~\ref{Figure:BTCUSD_market_impact_5s_resample}, left image), a positive relationship between trade size and price change can be observed. However, despite the generally positive relationship, some large discrepancies in impact between orders of similar sizes are observed. This is likely due to the differences in prevailing market liquidity and volatility mentioned previously. A common model to control for prevailing liquidity, current volatility, as well as the nature of the non-instantaneous market impact, is the aforementioned \textit{Square Root model} \cite{torre1997barra}. The general form of this model is shown in Equation~\ref{Equation:square_root_law}, where the current period market impact, $I$, is modelled as a function of the volume of the orders in the current resample period (5 seconds), $Q$, and the ''recent`` time horizon (1 hour), $T$. For control factors, $\sigma_T$ is the price volatility over time $T$, $V_T$ is the average traded volume over time $T$, $\delta$ the exponent of the diminishing market impact, usually assumed to be around 0.5 (hence the \textit{Square Root law}), and a constant, $k$.

\begin{equation}
\label{Equation:square_root_law}
I(Q, T) = k\sigma_T(\frac{Q}{V_T})^{\delta}
\end{equation}

We take an incremental approach to analyse the initial market impact. We start by ignoring the exponent, $\delta$, and adjust the market data by prevailing price volatility, $\sigma_T$, and the average volume, $V_T$. This yields a plot that appears better correlated (Figure~\ref{Figure:BTCUSD_market_impact_5s_resample}, right image). Note that we calculate $\sigma_T$ using the standard deviation of period-by-period price change over time $T$, and $V_T$ using a linearly increasing weighted average of orders over time $T$, where volume from the most recent period is assigned the highest weighting factor.

\begin{figure}[!htb]
  \centering
  \includegraphics[width=\textwidth]{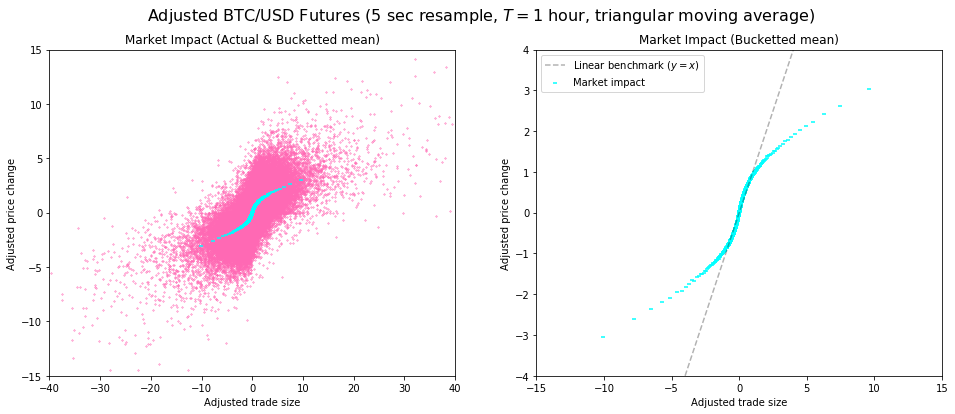}
  \caption{BTC/USD Futures Market Impact with percentile buckets (5 sec resample)}
  \label{Figure:BTCUSD_market_impact_5s_scatter_adjusted_buckets}
\end{figure}

Despite the pleasing aesthetics of the plot, this visualisation is slightly misleading, due to the high density of data points (over 800,000). Therefore, when the orders are gathered into buckets of order size and their mean market impacts plotted on top of the scatter plot (cyan lines, Figure~\ref{Figure:BTCUSD_market_impact_5s_scatter_adjusted_buckets}, left image), we find the relationship between order size and price change to be non-linear. Isolating and looking closer into the bucketed means (Figure \ref{Figure:BTCUSD_market_impact_5s_scatter_adjusted_buckets}, right image), increasing resistance of price change to traded volume can be observed clearly, suggesting that the exponent $\delta$ in Equation \ref{Equation:square_root_law} must have a value less than 1.

\begin{figure}[!htb]
  \centering
  \includegraphics[width=\textwidth]{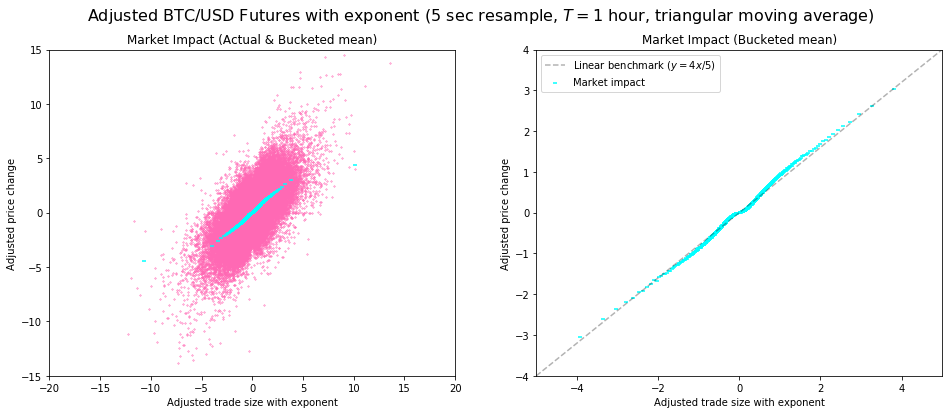}
  \caption{BTC/USD FuturesMarket Impact with percentile buckets (5 sec resample)}
  \label{Figure:BTCUSD_market_impact_adjusted_bucket_with_exponent}
\end{figure}

Using the data, we numerically find the value of $\delta$ using an optimisation routine. The results of the optimisation yield an outcome $\delta = 0.59$. When this exponent is applied to the data following Equation~\ref{Equation:square_root_law}, the data and the bucketed mean impact are transformed to that shown in Figure~\ref{Figure:BTCUSD_market_impact_adjusted_bucket_with_exponent}. Although the scatter plot exhibits a slight kink around zero net volume, it nonetheless displays the relationship between trade size and price change explained by Equation~\ref{Equation:square_root_law} linearly. We believe this empirical evidence is sufficient in showing the validity of the \textit{Square Root law} in the BTC/USD Futures market, and therefore adequate to be used as a baseline result to compare our simulated results.

Before concluding our evidence for initial market impact, we quickly note the slight non-linearity in the adjusted bucket plots in Figure~\ref{Figure:BTCUSD_market_impact_adjusted_bucket_with_exponent}. The explainability near zero net trade size is non-linear for two reasons. Firstly, small net orders tend to be traded in a differing liquidity environment on either side of the book. In other words, 100 units of buy orders followed by 100 units of sell orders will not necessarily bring the price back to the starting point, depending on the prevailing liquidity on the buy and sell side of the book. Second, the reversion of market impact from previous periods will account for a larger portion of the overall market impact near the zero mark (since the current market impact is small), distorting the relationship between current order size and market impact.

\subsection{Reversion in Market Impact}

\begin{figure}[!htb]
  \centering
  \includegraphics[width=\textwidth]{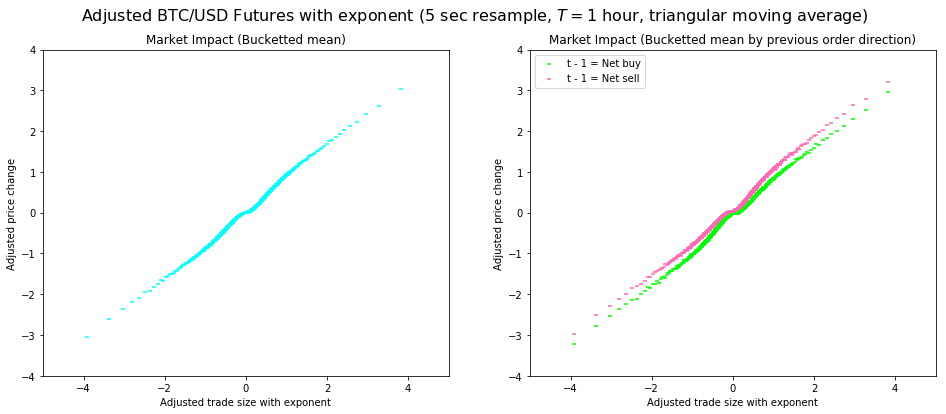}
  \caption{BTC/USD Futures market impact coefficient over time (5 sec resample, volume \& volatility adjusted, including exponent)}
  \label{Figure:BTCUSD_market_impact_5s_buckets_previous_sign}
\end{figure}

We now present evidence for reversion in market impact in two ways. The first method is by showing the intertemporal conditionality between the previous trade size and the current market impact. If there is no reversion in market impact, there should be no intertemporal conditionality between the previous trade size and the current market impact. Therefore, under this assumption, the sign of the previous period volume (net buy or net sell) should not influence the current market impact in any way. However, when the current market impact is divided into two groups based on the buy/sell status of the previous period, we observe a gap between the groups (Figure~\ref{Figure:BTCUSD_market_impact_5s_buckets_previous_sign}). This analysis indicates that periods following a previous period of net buy orders on average cause a lower market impact than those following a period of net sell orders. This means the previous order exerts market impact pressures in the opposite direction on the current period, evidencing reversion in impact.

The second approach looks beyond the first-term lag. As described by the propagator model \cite{gatheral2010no}, we expect the price impact of trade to revert in a decaying fashion, as the time between the trade and price impact observation increases. This follows Equation~\ref{Equation:propagator_model}, where the total market impact, $M(t)$, over period $t$ is the accumulation of all initial impacts made by trades across the period, $f(\dot{x_s})$, after the decay of these impacts have been taken into consideration following the decay kernel $G(t-s)$, plus a noise term $\epsilon$. Here, the $\dot{x_s}$ represents the rate of trading across our period $t$ and is assumed to be constant (thus we assume trades are made at a constant rate, and the integral of $\dot{x_s}$ over the period $t$ is equal to the sum of all trades across this period). We have shown in the previous section that the initial impact follows the functional form represented by the \textit{Square Root model}. Therefore function $f$ can be represented by function $I$ from Equation~\ref{Equation:square_root_law}, where the recent time horizon, $T$, is fixed to be 1 hour. This allows us to find out the functional form of the decay function, $G(t-s)$, in the BTC/USD Futures market in an empirical manner, by investing the market impact $M(t)$ for varying levels of $t$. Note that we ignore the noise term $\epsilon$ and assume it to be zero in our work.

\begin{equation}
\label{Equation:propagator_model}
M(t) = \int_{0}^{t} f(\dot{x_s})G(t-s) \,ds + \int_{0}^{t} \epsilon \,dZ_s
\end{equation}

\begin{figure}[!htb]
  \centering
  \includegraphics[width=\textwidth]{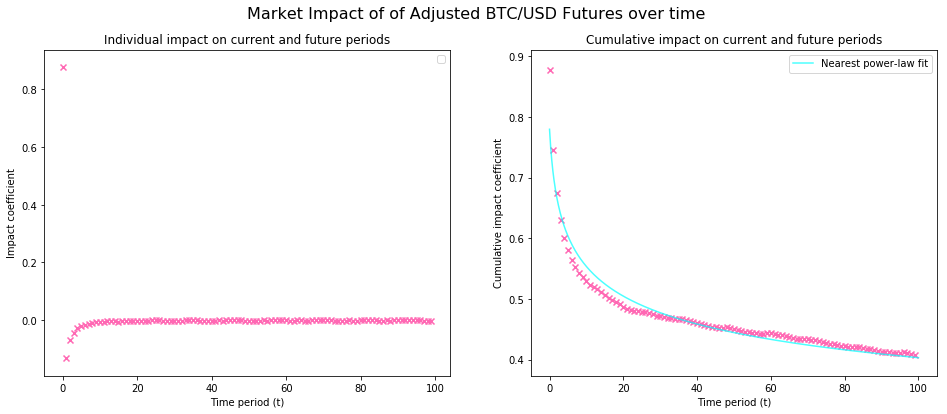}
  \caption{BTC/USD Futures market impact decay over time (5 sec resample, volume \& volatility adjusted, including exponent)}
  \label{Figure:BTCUSD_market_impact_adjusted_decay}
\end{figure}

By regressing the adjusted order size from time $t-100$ to $t$ against adjusted price change at time $t$ for all available $t$, the resulting coefficients show the average market impact of a unit trade on the present and future periods (Figure~\ref{Figure:BTCUSD_market_impact_adjusted_decay}). The coefficients show a large initial impact at $t=0$, followed by the reversion in impact, as exhibited by small negative coefficients at $t>0$. When the cumulative market impact of a unit order at time $t$ is plotted over 100 periods as before (Figure~\ref{Figure:BTCUSD_market_impact_adjusted_decay}), a decay in market impact following a power law can be observed.

\subsection{Meta Orders}

The final section of our analysis of the BTC/USD Futures market is regarding meta orders. In general, large orders are known to be broken down into many smaller orders in an attempt to hide the total transaction volume and reduce the market slippage of the transaction. This is done in various methods, including the well-known VWAP (Volume Weighted Average Price) and TWAP (Time Weighted Average Price) execution strategies. Breaking down a large order into smaller chunks leads to the idea of meta orders, which spawn consecutive orders in the same direction. This idea can be verified by checking the autocorrelation of order signs between consecutive orders. 

\begin{figure}[!htb]
  \centering
  \includegraphics[width=\textwidth]{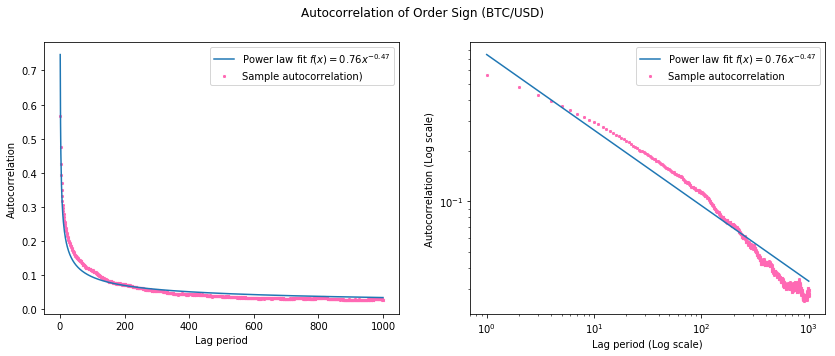}
  \caption{BTC/USD Autocorrelation of Order Sign (Actual Data, March 2021)}
  \label{Figure:BTCUSD_autocorrelation_of_order_sign_actual_march_2021}
\end{figure}

As shown in Figure~\ref{Figure:BTCUSD_autocorrelation_of_order_sign_actual_march_2021}, the BTC/USD Futures market exhibit a clean power-law decay in this autocorrelation of order signs, supporting the existence of these meta orders. This autocorrelation is also something that we look to include in our market simulation, in an attempt to produce the desired market reaction to a transaction.

\section{Simulated market impact}

In section~\ref{Section:market_impact_in_the_real_world}, we identified and analysed the market impact observed in the BTC/USD Futures exchange using the \textit{Square Root law} and the \textit{Propagator model}. This allowed us to divide up impact into two components: the initial impact and the decay kernel. Our objective now is to ensure that our market simulation replicates these observations as closely as possible. For this, we analyse an existing approach to modelling the limit order book and then create a novel approach to improve upon the shortcomings of the existing model.

\subsection{Santa-Fe Model}

Our starting point is the \textit{Santa Fe model} of the market \cite{farmer2005predictive}, originally developed by researchers at the Santa Fe institute. This model is created using stochastic processes to populate the limit order book following a zero-intelligence approach. Within the model, limit orders of fixed size arrive following a Poisson process across a predetermined price range. The arrival rate at each price is independent of one another and independent of the number of outstanding orders at each level. At the same time, each of these limit orders is cancelled following a separate Poisson process, which is again independent of the state. Simultaneously, market orders arrive following yet another Poisson process that takes out limit orders. In doing so, the LOB is populated as a function of limit order arrival \& cancellation rate, and market order arrival rate.

We replicate this model in the \textit{ABIDES} simulator by representing these processes using augmented Zero-Intelligence agents. This is done by dividing the Zero-Intelligence agents into limit order agents (Algorithm~\ref{santa_fe_zero_intelligence_pseudocode_limit}) and market order agents (Algorithm~\ref{santa_fe_zero_intelligence_pseudocode_market}). Although it is possible within the ABIDES platform to create a single type of Zero-Intelligence agent that is allocated a probability of behaving as a limit or a market order agent, this increases the run-time of the algorithm, without making any difference in the system behaviour. It may also cause the agents to sometimes transact against their own limit orders, which makes little sense from an explainability perspective.

\begin{algorithm}
 \caption{Modified Zero-Intelligence algorithm for Santa-Fe model (Limit order agent)}
 \label{santa_fe_zero_intelligence_pseudocode_limit}
    \SetAlgoLined
    
    $p_{cancel} := P$(Cancelling an existing limit order)\\
    Generate a random number $n\sim \mathcal{U} (0, 1)$ \\
    
    \eIf{$n < p_{cancel}$}{

        Cancel oldest limit order if one exists\\

    }{
     
     Obtain private valuation, $V \sim \mathcal{U} \{1, 100\}$\\ 
     
     Query best market ask, $A'$\\
     Query best market bid, $B'$\\
     Calculate mid price $M' = (A' + B')/2$ \\
     
     \eIf{$V < M'$}{
     
        Place limit order to buy at $V$
     
        }{
        
        Place limit order to sell at $V$
        
        }
    }

\end{algorithm}

\begin{algorithm}
 \caption{Modified Zero-Intelligence algorithm for Santa-Fe model (Market order agent)}
 \label{santa_fe_zero_intelligence_pseudocode_market}
    \SetAlgoLined

    Generate a random number $m \sim \mathcal{U} (0, 1)$\\
        
    \eIf{$m \leq 1/2$}{
        Place market order to buy
    }{
        Place market order to sell
    }

\end{algorithm}

\begin{figure}
\centering
\begin{minipage}{.5\textwidth}
  \centering
  \includegraphics[width=\linewidth]{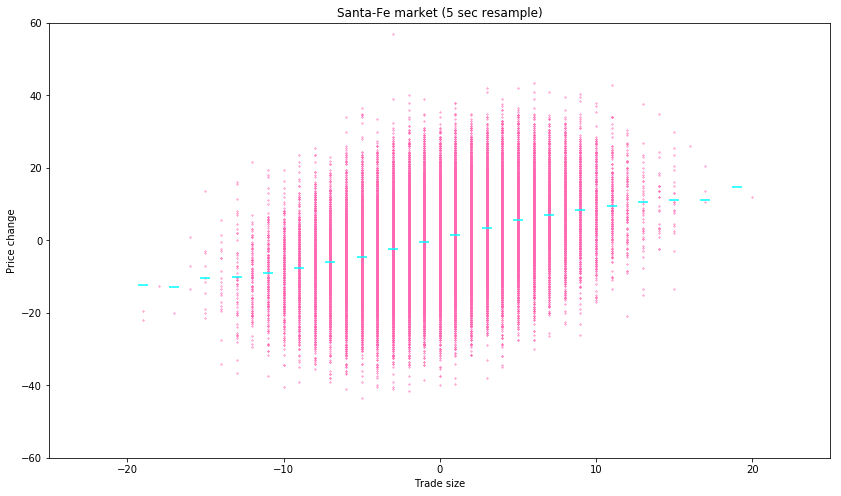}
  \caption{Santa-Fe Market Impact with percentile bucket average impact (5 sec resample)}
  \label{Figure:Santa_Fe_market_impact_5s_resample_scatter_bucket}
\end{minipage}%
\begin{minipage}{.5\textwidth}
  \centering
  \includegraphics[width=\linewidth]{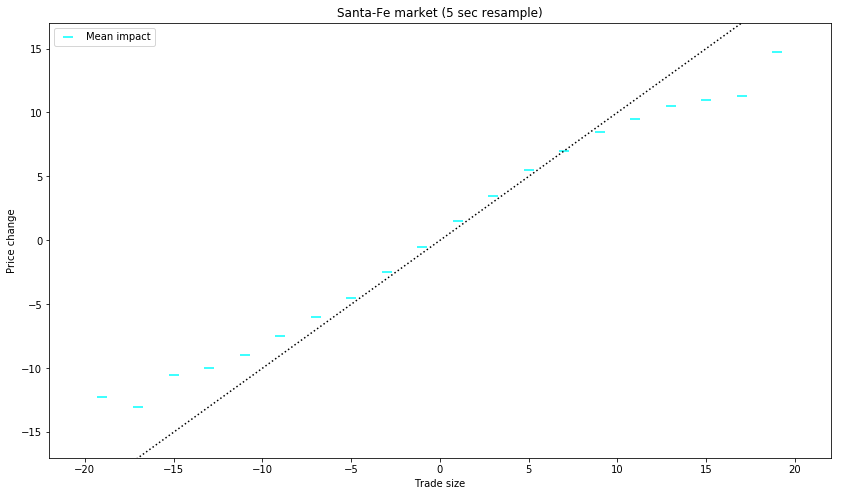}
  \caption{Santa-Fe Market Impact with percentile bucket average impact (1 sec resample)} \label{Figure:Santa_Fe_market_impact_5s_resample_bucket}
\end{minipage}
\end{figure}

\begin{figure}[!htb]
  \centering
  \includegraphics[width=0.5\textwidth]{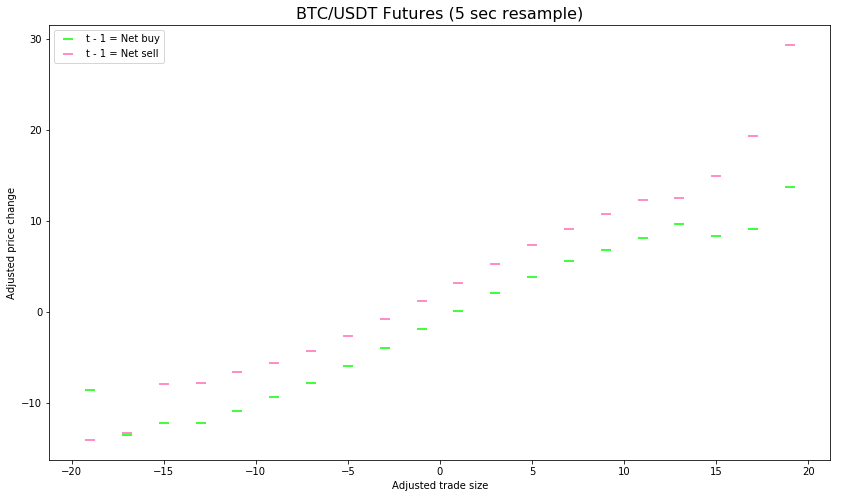}
  \caption{Santa-Fe Market Impact with percentile bucket average impact by previous order direction (5 sec resample)}
  \label{Figure:SantaFe_market_impact_5s_bucket_previous_sign}
\end{figure}

\begin{figure}[!htb]
  \centering
  \includegraphics[width=\textwidth]{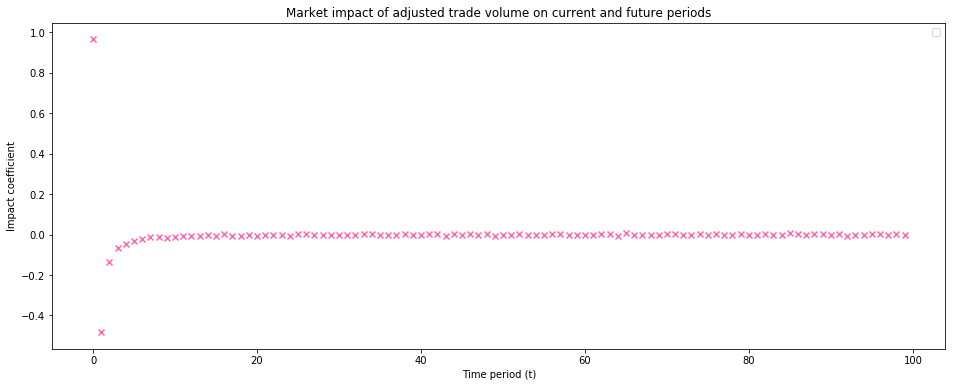}
  \caption{Santa-Fe market impact coefficient over time (5 sec resample)}
  \label{Figure:SantaFe_market_impact_5s_coef}
\end{figure}

\begin{figure}[!htb]
  \centering
  \includegraphics[width=\textwidth]{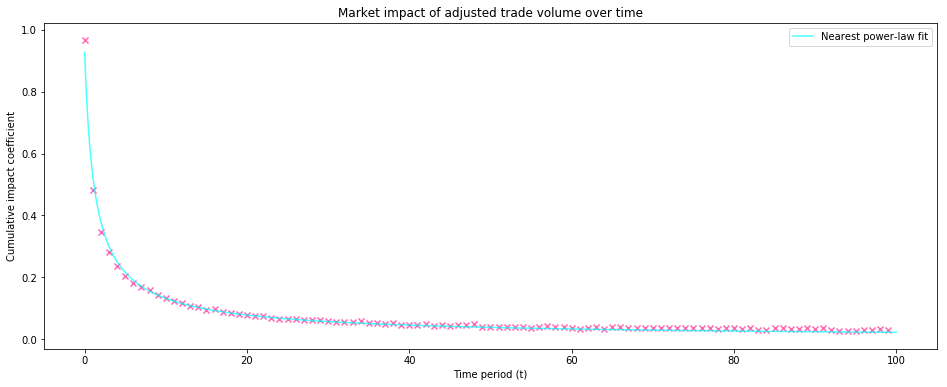}
  \caption{Santa-Fe market impact decay over time (5 sec resample)}
  \label{Figure:SantaFe_market_impact_5s_decay}
\end{figure}

\begin{figure}[!htb]
  \centering
  \includegraphics[width=\textwidth]{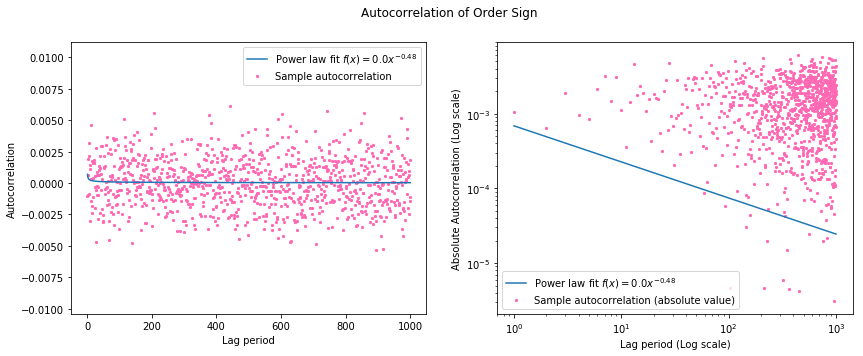}
  \caption{Autocorrelation of Order Sign (Santa-Fe Simulation)}
  \label{Figure:SantaFe_autocorrelation_of_order_sign}
\end{figure}

\begin{figure}[!htb]
  \centering
  \includegraphics[width=\textwidth]{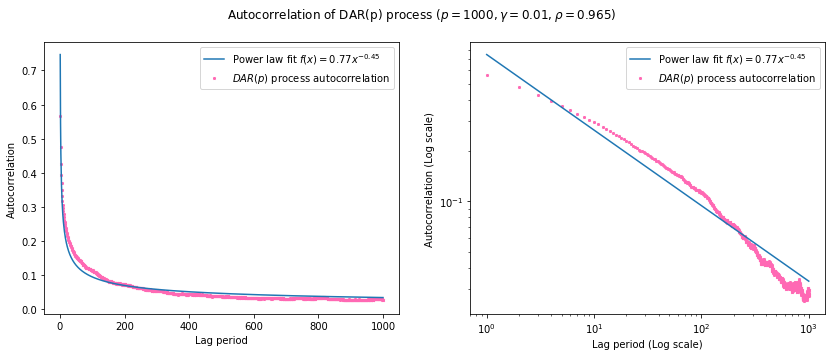}
  \caption{Santa-Fe autocorrelation of Order Sign (Simulated Data, March 2021)}
  \label{Figure:DAR_p_process_autocorrelation_tuned}
\end{figure}

Reproducing the Santa Fe model in the ABIDES simulator yields very good results on the market impact front. The diminishing initial market impact can be observed in Figure~\ref{Figure:Santa_Fe_market_impact_5s_resample_scatter_bucket} and Figure~\ref{Figure:Santa_Fe_market_impact_5s_resample_bucket}, and the reversion in market impact can also be observed in Figures~\ref{Figure:SantaFe_market_impact_5s_bucket_previous_sign}, \ref{Figure:SantaFe_market_impact_5s_coef} and \ref{Figure:SantaFe_market_impact_5s_decay}. However, it was unable to show desirable autocorrelation in order sign to produce the idea of meta orders (Figure~\ref{Figure:SantaFe_autocorrelation_of_order_sign}). To address this discrepancy, we use a method created by Taranto et al \cite{taranto2016linear} and incorporate a stochastic process called DAR(p), or the \textbf{D}iscrete \textbf{A}uto\textbf{R}egressive process of order $p$, to the market order agent (Algorithm~\ref{santa_fe_zero_intelligence_pseudocode_market_dar_p}). This additional augmentation maintains the behaviour of the initial market impact and its decay observed in the unaugmented Santa-Fe simulation, whilst now exhibiting order sign correlation that decays in a desirable fashion (Figure~\ref{Figure:DAR_p_process_autocorrelation_tuned}). Note that we tune the parameters of the DAR(p) process in Algorithm~\ref{santa_fe_zero_intelligence_pseudocode_market_dar_p} by using a Monte-Carlo method to minimise the difference between the parameters of the power law fit observed in the real world (Figure~\ref{Figure:BTCUSD_autocorrelation_of_order_sign_actual_march_2021}) and the one observed in the simulation (Figure~\ref{Figure:DAR_p_process_autocorrelation_tuned}).

The results obtained from the \textit{Santa Fe model} are very promising. However, it doesn't quite provide us with the functionality fully equipped for real-world simulation. Firstly, it lacks model explainability. The orders submitted in a stochastic process by Zero-Intelligence agents do not represent a type of market participant. Instead, it assumes that aggregate participation in the market follows a Poisson process. This hampers our ability to attribute market behaviour to a specific type of participant or test the impact of adjusting agent-specific behaviour on the overall market. Secondly, the model relies on this aggregated behaviour to remain in equilibrium to exhibit desirable market impact. This means that adding additional participants to the simulation may destabilise this equilibrium and push the simulation away from desirable behaviour. Both these shortfalls remove some of the key advantages of using an agent-based model. Another limitation of the Santa Fe model is its inability to follow an underlying price series. As both limit and market orders arrive independently around the mid-price, the equilibrium either mean-reverts or trends away from the initial equilibrium on its own accord, based on the system's lag-dependent volatility \cite{BOUCHAUD2018393}. This is why we opt to develop a new model that addresses these shortcomings.

\begin{algorithm}
 \caption{Modified Zero-Intelligence algorithm for Santa-Fe model (Market order agent, DAR(p) process)}
 \label{santa_fe_zero_intelligence_pseudocode_market_dar_p}
    \SetAlgoLined

    Generate a series of n random numbers $m_i \sim \mathcal{U} (0, 1) * $ to form array $M = (m_1, m_2, m_3, ..., m_n)$\\

    Generate a ``parent'' index $\ell \leq n$ from $\lambda_\ell$

    where $\lambda_\ell \sim \ell^{(\gamma-3)/2}$\\

    \eIf{$p \leq r \sim \mathcal {U}_{[0,1]}$}{
    
        $m_{t} = m_\ell$

    }{

        $m_{t} = 1-m_\ell$

    }

    \eIf{$m_{t} = 1$}{
    
        place buy market order

    }{

        place sell market order

    }

    shift index of M by 1

    $(m_1, m_2, m_3, ..., m_n) \rightarrow (m_2, m_3, m_4, ..., m_{n+1})$

    remove the last element $m_{n+1}$ \\
    
    add $m_t$ as the first element $m_1$ \\

\end{algorithm}

\subsection{PRIME}

To address the issues with \textit{Santa Fe model}, we propose a novel simulation methodology we call \textit{PRIME}, or the \textbf{P}rice-\textbf{R}everting \textbf{I}mpact \textbf{M}odel of a cryptocurrency \textbf{E}xchange. This model is able to follow an underlying price series and replicate the desired market microstructural properties of the cryptocurrency exchange, whilst maintaining the interactivity and the explainability of the simulation. We achieve this by augmenting and adding to the existing \textit{Santa Fe model} in various ways. 

Firstly, the limit order agents no longer ``receive'' a private valuation from a predetermined price range. Instead, they calculate their private valuation by querying the prevailing mid-price of the simulation and adding a small perturbation drawn from a uniform distribution. Secondly, at the beginning of the simulation, the LOB is populated in a linearly increasing manner around a specified starting price. This allows the limit order agents to query the mid-price and populate the limit order book from the start. Lastly, market order agents have access to the true underlying price series via the oracle, again with uniformly distributed perturbation. These changes make the limit-order agents function similarly to background traders, and market-order agents become the true fundamental agents who exclusively execute their strategies aggressively. Lastly, we include technical agents that either trend-follow or mean-revert in the market by submitting market orders into the system. Following these changes, the agent configuration we use in the PRIME model is 1000 Zero-intelligence limit order agents, 30 Zero-intelligence market order agents, 10 trend-following agents and 10 mean-reversion agents.

\begin{figure}[!htb]
  \centering
  \includegraphics[width=\textwidth]{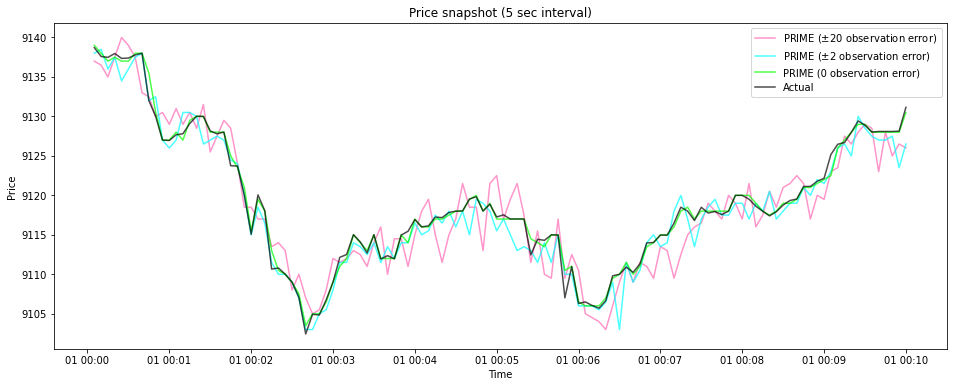}
  \caption{PRIME Price Series with various observation error (5 sec resample)}
  \label{Figure:PRIME_price_series_with_error}
\end{figure}

The \textit{PRIME} model behaves very similarly to the \textit{Santa Fe model} if the underlying value of the asset is fixed. In this case, limit orders arrive independently following a Poisson process on both sides of the prevailing mid-price, with an independent cancellation rate attributed to each order. Market orders also arrive following a Poisson process, and if the prevailing market equilibrium is equal to the underlying price series, the model remains identical to the \textit{Santa-Fe model}. However, if the model mid-price is not equal to the true price, the fundamental valuation received by market-order agents will be skewed away from the simulation's mid-price. This results in the market orders arriving in an asymmetric manner (different buy/sell order arrival rate) to push the model equilibrium back towards the true price. This forces \textit{PRIME} to be always mean-reverting, which is not the case with the Santa-Fe model. Conversely, if the true price of the underlying asset is not fixed at the initial price, and instead fluctuates following a predetermined path (e.g. historical price series), then \textit{PRIME} offers a simulation that follows the underlying price series (Figure~\ref{Figure:PRIME_price_series_with_error}), whilst largely maintaining the market mechanics of the Santa-Fe model. In addition, thanks to the market-order agents in \textit{PRIME} exerting pressure on the system to revert to the true price, the simulation can effectively interact with other agents, as perturbations made to the system will be equilibrated based on the true market price observed by fundamental agents. This makes \textit{PRIME} better equipped to explain and represent real-world markets.

\begin{figure}[!htb]
  \centering
  \includegraphics[width=\textwidth]{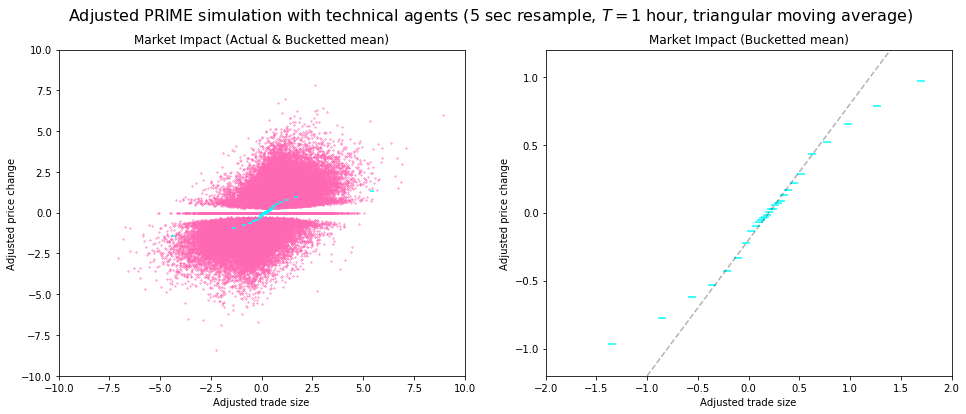}
  \caption{PRIME Market Impact with percentile buckets (5 sec resample)}
  \label{Figure:PRIME_market_impact_5s_scatter_adjusted_buckets}
\end{figure}

\begin{figure}[!htb]
  \centering
  \includegraphics[width=\textwidth]{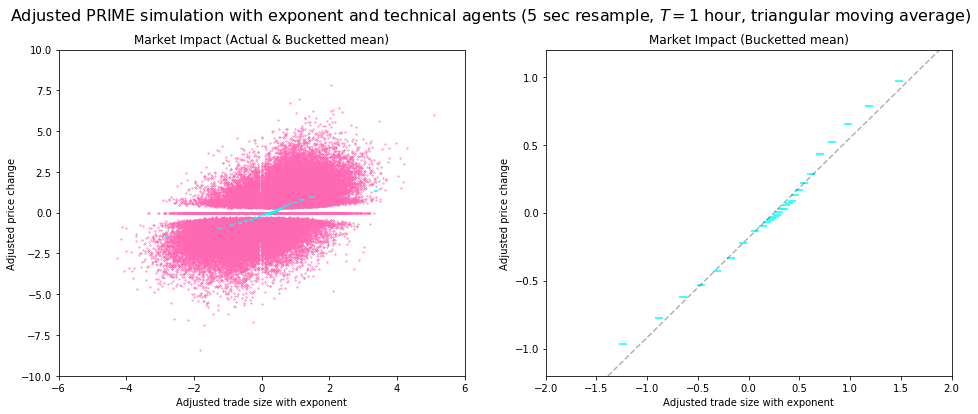}
  \caption{PRIME Market Impact with percentile buckets (5 sec resample)}
  \label{Figure:PRIME_market_impact_adjusted_bucket_with_exponent}
\end{figure}

We assess \textit{PRIME} using the same methodology as before. Firstly, we seek evidence of the \textit{Square Root law} by looking at the adjusted impact plot (Figure~\ref{Figure:PRIME_market_impact_5s_scatter_adjusted_buckets}). As before, we initially apply only volume and volatility adjustment without the exponent, $\delta$ (See Equation~\ref{Equation:square_root_law}). Although there is a disjoint in adjusted returns, due to the limited resolution of the price levels in the \textit{PRIME} model (as per Santa-Fe, limit orders are only submitted within $\pm50$ ticks of the mid-price), the resulting output clearly shows a diminishing impact as trade size increases. Secondly, when an exponent of value 0.74 is applied to the data, the relationship becomes much more linear, as shown by the cyan bucketed mean impact plots in Figure~\ref{Figure:PRIME_market_impact_adjusted_bucket_with_exponent}.

\begin{figure}[!htb]
  \centering
  \includegraphics[width=\textwidth]{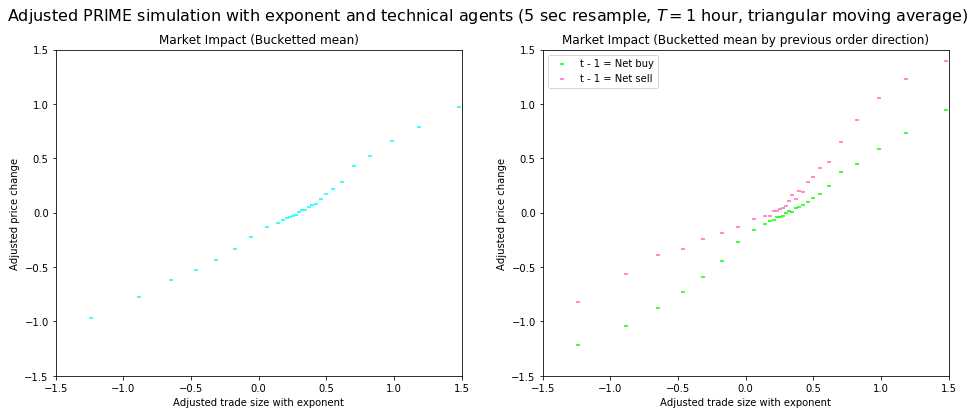}
  \caption{PRIME market impact coefficient over time (5 sec resample, volume \& volatility adjusted, including exponent)}
  \label{Figure:PRIME_market_impact_5s_buckets_previous_sign}
\end{figure}

\begin{figure}[!htb]
  \centering
  \includegraphics[width=\textwidth]{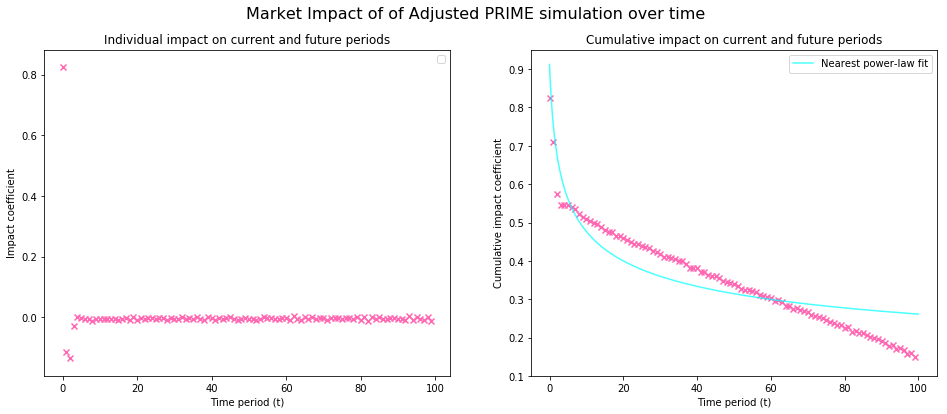}
  \caption{PRIME market impact decay over time (5 sec resample, volume \& volatility adjusted, including exponent)}
  \label{Figure:PRIME_market_impact_5s_adjusted_decay}
\end{figure}

Next, we look for evidence of reversion by dividing up orders by previous period order sign (Figure~\ref{Figure:PRIME_market_impact_5s_buckets_previous_sign}, right image). The market impact following a buy order clearly shows a lower market impact than those following a sell order, in line with the behaviour observed in the real world (Figure~\ref{Figure:BTCUSD_market_impact_5s_buckets_previous_sign}). In addition, evidence of reversion is also found when looking at the intertemporal impact coefficients and their cumulative decay in Figure~\ref{Figure:PRIME_market_impact_5s_adjusted_decay}. While there is a reversion in impact that decays over time, the shape of the decay is very linear after the first few terms, as opposed to the diminishing decay observed in the real world (Figure~\ref{Figure:BTCUSD_market_impact_adjusted_decay}).

\begin{figure}[!htb]
  \centering
  \includegraphics[width=\textwidth]{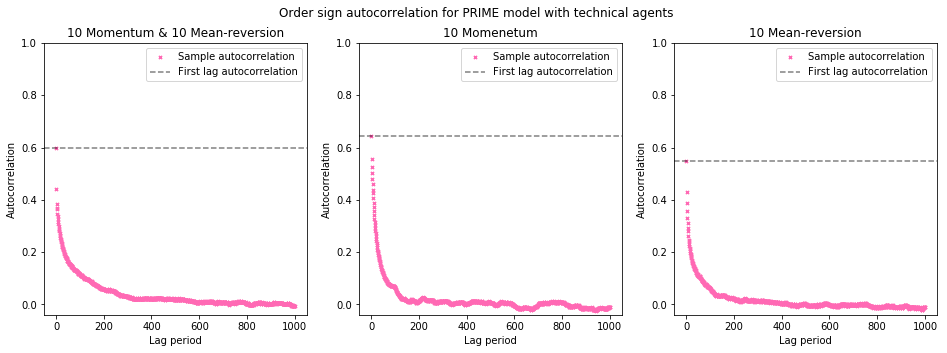}
  \caption{Santa-Fe autocorrelation of Order Sign (Simulated Data, March 2021)}
  \label{Figure:PRIME_ordersign_autocorr_50_0_0_vs_40_10_0_vs_40_0_10}
\end{figure}

Lastly, we see a promising result regarding the autocorrelation of order sign in the left image of Figure~\ref{Figure:PRIME_ordersign_autocorr_50_0_0_vs_40_10_0_vs_40_0_10}. Like the real-world data and the augmented \textit{Santa-Fe model}, we observe a decay that roughly follows a power law. This is especially encouraging, as the \textit{PRIME} model does not embed artificial stochastic processes into the participating agents to generate this result. In fact, the autocorrelation occurs from the technical agents within the PRIME model. The momentum agents who place orders in the same direction as previous orders increase the order sign autocorrelation (Figure~\ref{Figure:PRIME_ordersign_autocorr_50_0_0_vs_40_10_0_vs_40_0_10}, middle image), whereas the mean reversion agents do the opposite and reduce the autocorrelation (Figure~\ref{Figure:PRIME_ordersign_autocorr_50_0_0_vs_40_10_0_vs_40_0_10}, right image).

\subsection{Model Evaluation}

Our analysis shows that both the \textit{Santa-Fe model} and the \textit{PRIME} model are able to replicate the three market microstructural properties we analysed in the BTC/USD Futures market: initial impact, reversion of impact, and the autocorrelation of order signs. Whilst the \textit{Santa-Fe model} exhibited cleaner and more desirable results regarding the reversion of impact and the autocorrelation of order signs, the model lacks fundamental bottom-up explainability, especially after adding in the DAR(p) process to the market order agent. In addition, the \textit{Santa Fe model} lacks the ability to follow an underlying price series, which is critical in replicating and comparing real-world markets to the simulation. On the other hand, the \textit{PRIME} model offers a novel framework that addresses these weaknesses, at the cost of slight deviation from the desired shape of impact reversion and the decay in order sign autocorrelation. The increased explainability of the \textit{PRIME} model as well as its ability to trace an underlying price series makes the model exceptionally useful as a benchmark agent-based simulation of the market, that can be used to test and train market strategies and offers researchers with a first bottom-up agent-based simulation that replicates the market microstructure of the BTC/USD Futures market.

\section{Related Work}

This work is an extension of many previous efforts to replicate a live market environment. McGroaty et al \cite{mcgroarty2019high} provide a good summary of past studies and divide them into four categories: theoretical equilibrium models; statistical order book models; stochastic models; and agent-based models. In addition to these models, there has been recent progress to represent the market and the Limit Order Book dynamics by leveraging Machine-Learning techniques.

Our attempt approaches this problem using the agent-based approach to incorporate as much explainability as possible using a bottom-up methodology. The ability to include and customise market participants in an intuitive manner is a key advantage of the agent-based model, and it has been increasing in popularity \cite{vyetrenko2019get, cho2021bit, coletta2022learning, wang2017spoofing} as computational resources became more abundant over the past two decades.

While the other methods of modelling the LOB were considered alongside the Agent-Based model, we found them to be unsuited for our work. The Equilibrium approach yields results that are too fragile in equilibrium due to the number of assumptions required to allow the model to equilibrate. The statistical and machine-learning approaches provide promising results \cite{bouchaud2002statistical, dixon2018sequence, zhang2019deeplob}, but they lack explainability, due to their focus on the model fit, making intuitive adjustments difficult to add in the market. Similarly, while stochastic models \cite{abergel2013mathematical, farmer2005predictive} offer more than the statistical models in terms of explainability, they still have the tendency to group market participants into large categories, and thus offer less explainability when compared to the agent-based method.

In addition, our analysis of the behaviour of the BTC/USD Futures market is an update on the previous work done by Donier et al \cite{donier2015million}. Their work analysed the impact on the BTC/USD market, with a focus on meta orders, before the rise in popularity and size of the crypto market. Our findings provide an update on the behaviour of the cryptocurrency market post the 2016 rally using publically available data, with an additional focus on the shape of the impact decay.

\section{Conclusion \& Future Work}

In this paper, we provide two contributions to the field of cryptocurrency and market simulations. We first identified and evidenced market microstructure in the BTC/USD Futures market using publically available data, by leveraging previous theories such as the \textit{Square root law} and the \textit{Propagator model}. Secondly, and more importantly, we developed a novel simulation framework that addresses the shortcoming of existing models whilst maintaining model explainability, by replicating the key microstructural properties of the crypto market using heterogeneous agents. This simulation framework can be used by researchers in various ways, from investors testing their trading strategies, regulators looking at market fragility risks, and academics benchmarking and comparing novel models which incorporate state-of-the-art algorithms.

In our future work, we will look to make improvements in a few areas. First, we would like to improve the shape of the impact reversion to match that seen in real-world markets. Secondly, we would like to empirically tune the \textit{PRIME} simulation against observed microstructure in the BTC/USD Futures market, by varying the ratio and the aggressiveness of participating agents. Thirdly, we would like to incorporate the idea of latent liquidity, by augmenting traders to incorporate a more complex execution strategy described by algorithms such as Zero-Intelligence Plus \cite{cli1997minimal}. Lastly, we would like to run an extensive study on the stability of our simulation, by subjecting the simulation to a wide range of external agents that interact with the environment using diverse strategies. Doing this will improve the accuracy of \textit{PRIME}, and will allow researchers to utilise the simulation with an additional layer of confidence.

\section*{Acknowledgments}
This work was supported by EPSRC Centre for Doctoral Training in Next Generation Computational Modelling (EP/L015382/1). We also acknowledge the use of the IRIDIS High Performance Computing Facility, and associated support services at the University of Southampton, in the completion of this work.

\bibliographystyle{unsrt}  
\bibliography{references}

\end{document}